\documentclass[12pt,floats]{article}
\usepackage{epsfig}
\begin{document}

\title{\bf Single particle spectra from information theory point of
           view} 
\author{F.S. Navarra$^{1}$, O.V. Utyuzh$^{2}$,  G. Wilk$^{3}$ and  
Z. W\l odarczyk$^{4}$  
\\
$^{1}${\it Instituto de F\'{\i}sica, Universidade de S\~{a}o Paulo,}\\
      {\it Caixa Postal 66318, 05389-970 S\~ao Paulo, Brazil,}\\
      {\it e-mail: navarra@if.usp.br}\\
$^{2}${\it The Andrzej So\l tan Institute for Nuclear Studies,}\\
      {\it Ho\.za 69, 00681 Warsaw, Poland,} \\
      {\it e-mail: utyuzh@fuw.edu.pl }\\[1ex]
$^{3}${\it Institute of Physics, \'Swi\c{e}tokrzyska Academy,} \\
      {\it \'Swi\c{e}tokrzyska 15, 25-405 Kielce, Poland,} \\
      {\it e-mail: wlod@pu.kielce.pl }}

\date{\today}
\maketitle

\begin{abstract}
It is demonstrated how to obtain the least biased description of
the single particle spectra measured in all multiparticle production
processes by using information theory approach (known also as MaxEnt
approach). The case of $e^+e^-$ annihilation in hadrons process is
discussed in more detail as an example. Comparison between MaxEnt
approach and simple dynamical model based on the cascade process is
presented as well.

\end{abstract}

\newpage

Information theory is nowdays widely used in all branches of sciences
to provide the {\it least biased, most probable} description of data
for which we have only limited amount of {\it information} \cite{Info}.
The information is defined by means of information entropy, which, in
turn, can assume extensive (Shannon) or nonextensive (Tsallis) form,
depending on circumstances (see \cite{Info,T,MaxEnt,qMaxEnt} for
details and further references). In our case where we are interested
in single particle rapidity distributions, $dN/dy$, maximization of
such entropy (known as MaxEnt procedure \cite{Info,MaxEnt,qMaxEnt})
with constraint given by the energy conservation leads to the
following form of the probability distribution to find one of the $N$
produced particles of transverse mass $\mu_T = \sqrt{\mu^2 + \langle
p_T\rangle^2}$ in the longitudinal phase space described by rapidity
$y$ such that energy of this particle is $E=\mu_T \cosh y$ and its
longitudinal momentum is $p_L=\mu_T \sinh y$: 
\begin{equation}
p(y)\, =\, \frac{1}{N}\frac{dN}{dy}\,
=\, \frac{1}{Z_q} \exp_q         \left( - \beta_q \cdot \mu_T \cosh y
\right). \label{eq:py} 
\end{equation}
This form is identical with that used in statistical models but now
$Z_q$ and $\beta_q$ are no longer {\it free parameters} to be fitted
when comparing with experimental data but instead are given by the
normalization condition and energy conservation
constraint\footnote{~~Here $\exp_q\left( x/\Lambda\right) = \left[1 +
(1-q)x/\Lambda\right]^{1/(1-q)} \stackrel{q \rightarrow
1}{\Rightarrow} \exp (x/\Lambda)$).},
\begin{equation}
\int_{-Y_m}^{Y_m}\, dy\, p(y)\, =\, 1 \qquad {\rm and}\qquad 
\int_{-Y_m}^{Y_m} \, dy\, \mu_T\cdot \cosh y \cdot [p(y)]^q\, =\,
\frac{\kappa_q \cdot W}{N} \label{eq:energy}
\end{equation}
(where $\pm Y_m$ are maximal rapidities available in rest frame of
hadronizing source, see \cite{MaxEnt,qMaxEnt} for details). As is
demonstrated in \cite{qMaxEnt,qMaxEnt1} with $q>1$ (responsible for 
dynamical fluctuations, see \cite{WW1,WW2} and $q$-inelasticity 
$\kappa_q$ (connected to true inelasticity $K_q$ defining a fraction
of the original available energy $W$ used for the production of
secondaries, $K_q = \kappa_q/(3-2q)$, cf. \cite{qMaxEnt}) as the only
parameters we can describe all data for  $pp$ and $p\bar{p}$
collisions as well as recent data for nuclear collisions. Here we
show in Fig. 1 that we can also fit fairly well $e^+e^-$ annihilation
data for which, by definition, $K_q=1$ (the whole energy must be used
for the production, there are no leading particles) and $q$ remains
the only free parameter. It turns out that in this case $q<1$
\footnote{To be contrasted with $q>1$ values needed to describe the
$p_T$ distributions in the same processes instead \cite{Bed}.}. This
fact indicates (see \cite{WW2}) that in $e^+e^-$ processes (where
$K_q=1$) one cannot obtain equilibrium state because the
corresponding temperature parameter depends now on energy: $T =
1/\beta_q = T_0 - (1-q)E$ and system is highly influenced by the
conservation of energy constraint. In Fig. 1 we have also shown
attempts to fit the same data using the well defined sequential decay
(cascading) hadronization developed in \cite{CAS} instead of
instantaneous hadronization\footnote{Cascade could be regarded as a
viable alternative because of the {\it a priori} cascade character of
the quark$\rightarrow$ gluon and gluon $\rightarrow$ quark-antiquark
processes preceding hadronization  expected to be present in $e^+e^-$
annihilations \cite{Data}.}. Cascade here is defined by the sequence
of decay processes, $M_l \rightarrow [M^{(1)}_{l+1} = k_1 M_l] +
[M^{(2)}_{l+1}=k_2 M_l]$ (proceeding until $M_l > 2\mu_T$) in which
all dynamics is summarily described by distribution $P(k)$ of decay
parameters $k_i \in (0,1)$ (such that in each vertex $k_1 + k_2 < 1$,
see \cite{CAS} for details). It is easy to realise that in this case
$dN/dy$ is given by a kind of random walk in rapidity space, which,
in turn, results in its gaussian-like shape (cf. Fig. 1). However,
one cannot find any reasonable values of decay parameters $k_i$ to
fit data with such distribution, whereas MaxEnt in its nonextensive
version can do it with $q=0.6$. The remaining small discrepancies for
small and large rapidities indicate therefore a need for some
additional information to be incorporated here (see, for example,
discussion in \cite{qMaxEnt1}). When introduced it could then be
again checked against experimental data. In this way we could always
obtain good description of data with only {\it minimal} number of
assumptions, i.e., with minimal information content
\cite{Info,MaxEnt}. This is the main advantage of the method
presented here and in our opinion it deserves further detailed
investigations by using wide spectrum of the available multiparticle
production data of all kinds. 

\begin{center}
\begin{tabular}{|cc|}
\hline
\begin{minipage}{6cm}
\begin{center}
\includegraphics[height=4.5cm,width=6cm]{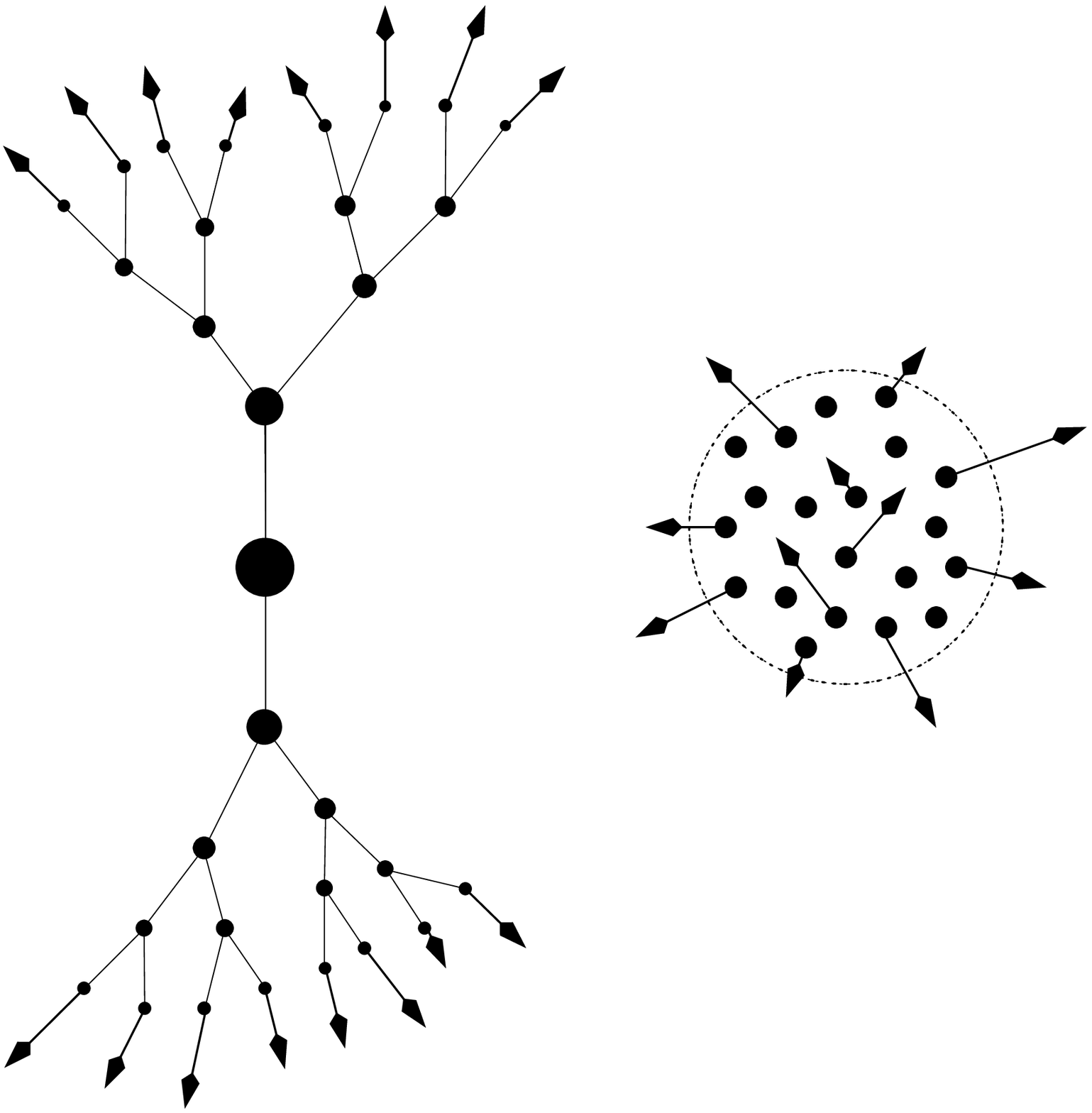} 
\end{center}
\end{minipage} &
\begin{minipage}{6cm}
\begin{center}
\includegraphics[height=4.5cm,width=6cm]{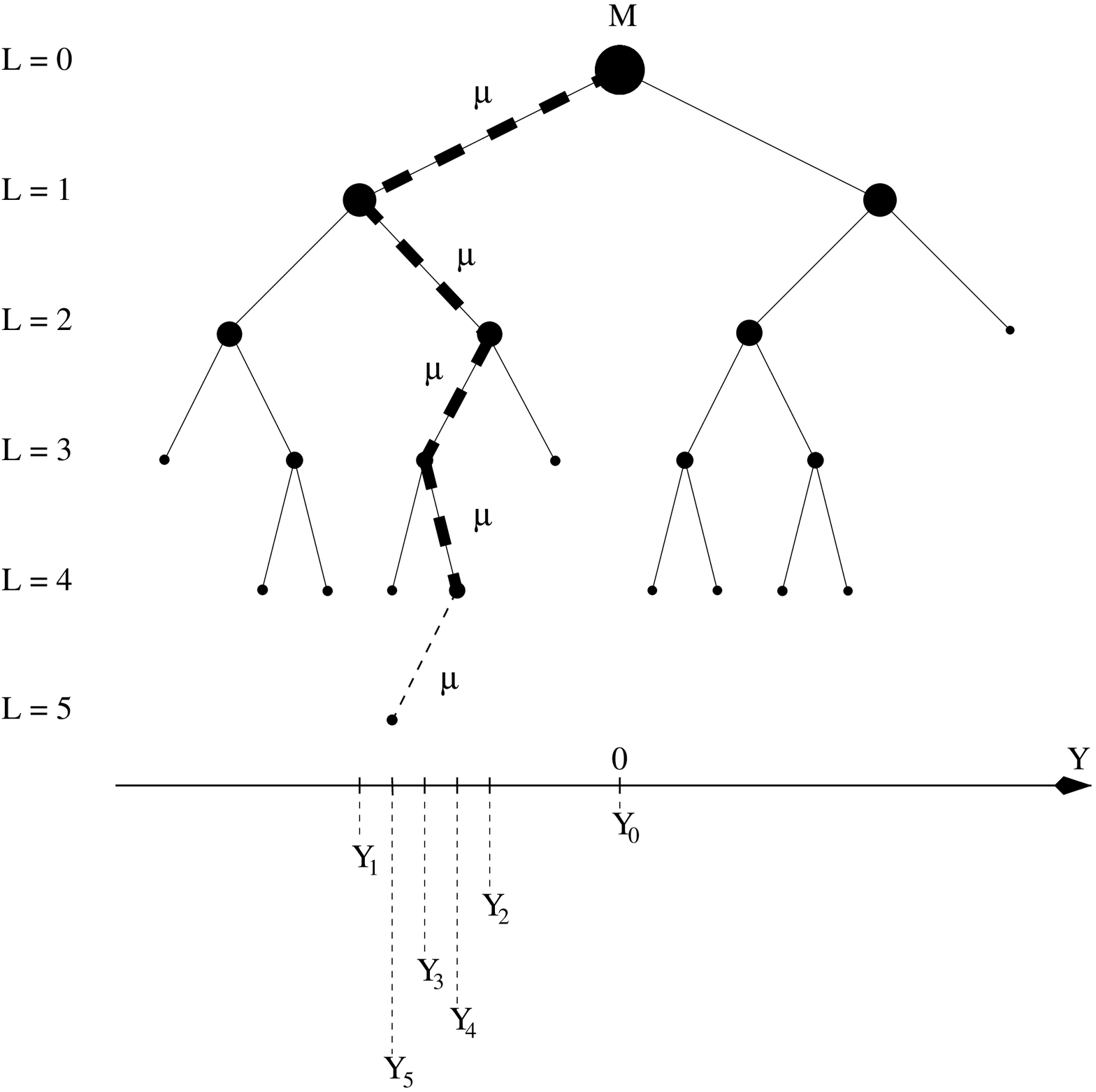}
\end{center}
\end{minipage} \\
\begin{minipage}{6cm}
\begin{center}
\includegraphics[height=5.5cm,width=6cm]{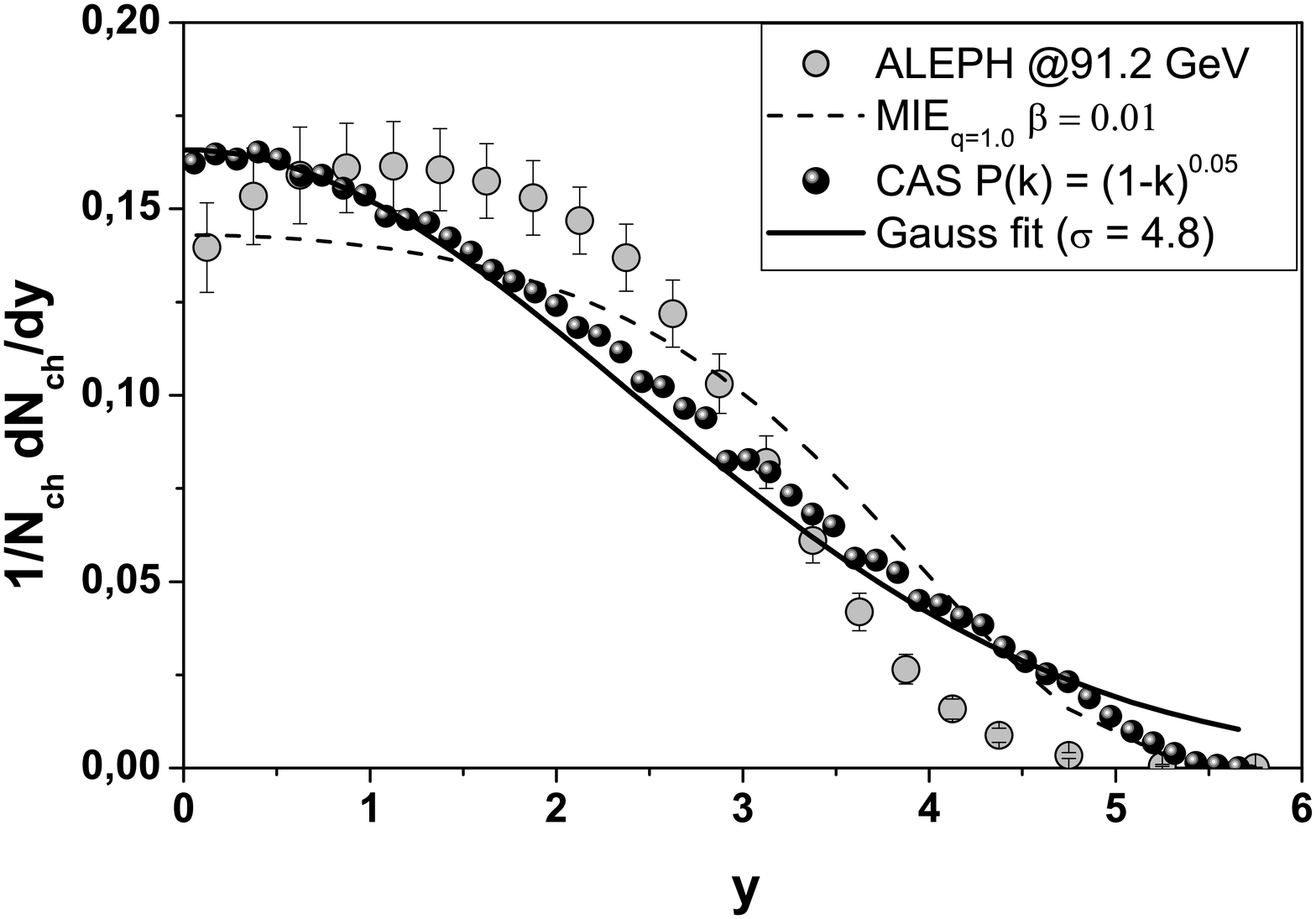}
\end{center}
\end{minipage} &
\begin{minipage}{6cm}
\begin{center}
\includegraphics[height=5.5cm,width=6cm]{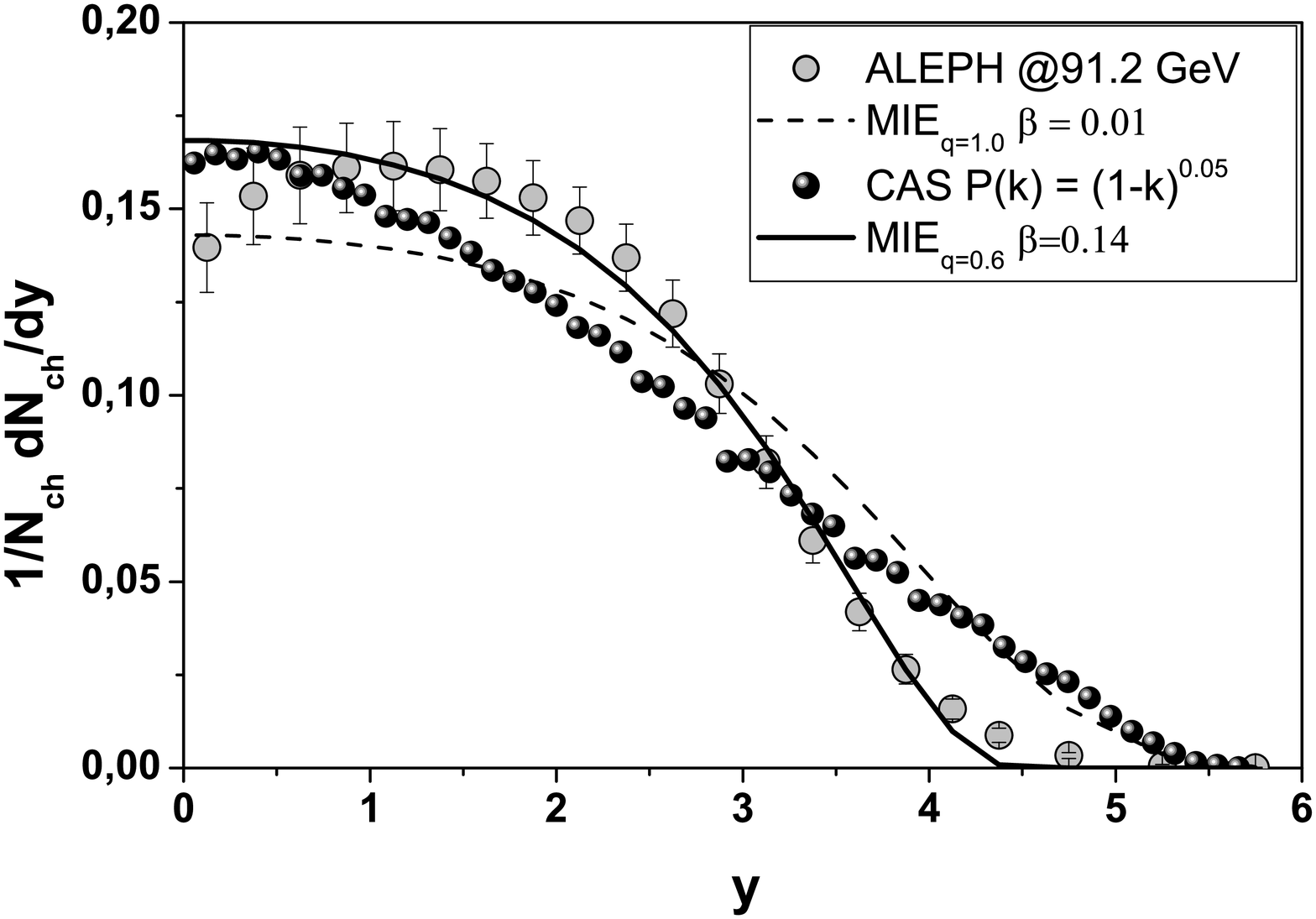}
\end{center}
\end{minipage} \\
\hline
\end{tabular} \\ 
\vspace{0.5cm}{\scriptsize Fig.1 Upper-left panel: visualization of
differences between sequential cascade hadronization (CAS) \cite{CAS}
and instantaneous one described by MaxEnt \cite{MaxEnt}. Upper-right
panel: closer look at the way $dN/dy$ arises in CAS (resembling
random walk in $y$-space). Lower panels: results of calculations of
$dN/dy$ using CAS and MaxEnt and compared with $e^+e^-$ annihilation
data by ALEPH \cite{Data}. Left panel -  comparison with CAS model
giving the same multiplicity \cite{CAS} (it can be extremely well
approximated by gaussian fit) and with MaxEnt for $K_q=1$ and $q=1$
(it does not fit data). Right panel - gaussian fit has been replaced
by MaxEnt fit with total inelasticity $K_q=1$ and with varying $q$;
for $q=0.6$ we obtain quite good agreement with data. 
 }
\end{center}

\end{document}